# Retracting fronts induce spatio-temporal intermittency

Pierre Coullet[1]* and Lorenz Kramer[2]
[1] *Institut Non Linéaire de Nice Sophia Antipolis, UMR 6618 CNRS, 1361 route des Lucioles, F-06560 Valbonne, France*
[2] *Physikalisches Institut, Universität Bayreuth, D-95440 Bayreuth, Germany*

(Dated: February 25, 2002)

The intermittent route to spatio-temporal complexity is analyzed in a simple model which displays a subcritical bifurcation without hysteresis. A new type of spatio-temporal complexity is found, induced by fronts which "convectively clean" the perturbations around an unstable state. The mechanism which leads to the existence of these "retracting fronts" is analyzed in the frame of simple models.

The intermittent route to spatio-temporal complex behaviors has received a great deal of attention during the last years, both from the theoretical and the experimental side [1]. In this letter we intend to propose one of the possible mechanisms for intermittency. This work has been initially motivated by an experimental result reported to us prior to publication [2] . At low Prandtl number convection with rotation arizes as a subcritical bifurcation. The experiments confirmed very precisely the value of the Raleigh number at the onset of convection and showed an apparent abrupt transition to intermittend convection which surprisingly did not exhibit hysteresis. The theoretical analysis demonstrates that the convective upper branch is unstable with respect to the Küppers-Lortz instability.

As a first attempt to understand this phenomenon, we consider the following model, which incorporates the subcritical character of the bifurcation and the instability of the upper branch as the keys elements.

$$\ddot{u} + \nu(u)\dot{u} + \frac{\partial V}{\partial u} = u_{xx} + \dot{u}_{xx} \qquad (1)$$

where $\nu(u) = 1 - \alpha u^2 + u^4$ and $V = -\lambda u^2/2 - u^4/4 + u^6/6$. It can be interpreted as a spatially extended version of the van der Pol model displaying a subcritical inverted pitchfork bifurcation at $\lambda = 0$ . The upper nonlinear branch is unstable with respect to a Andronov-Hopf bifurcation for an appropriate choice of $\alpha > \alpha_H = \frac{3+2\lambda+\sqrt{1+4\lambda}}{1+\sqrt{1+4\lambda}}$. Solutions of the homogeneous system, *i.e.* when spatial derivatives are dropped out from Eq. (1), and their bifurcations can be described in great detail. When $\alpha > 2.5$, the nonlinear branch appears unstably. In the following $\alpha$ is set to a fixed value ($\alpha = 3.8$) and $\lambda$ is varied. Then, in the homogeneous system, one has an abrupt transition to a large amplitude van der Pol like limit cycle which is for $\lambda > 0$ the global attractor. This limit cycle persists for negative values of $\lambda$ and hysteresis is observed.

In the spatially extended system the large amplitude limit cycle, although it is still strongly stable, is no longer the global

*Professor at the Institut Universitaire de France

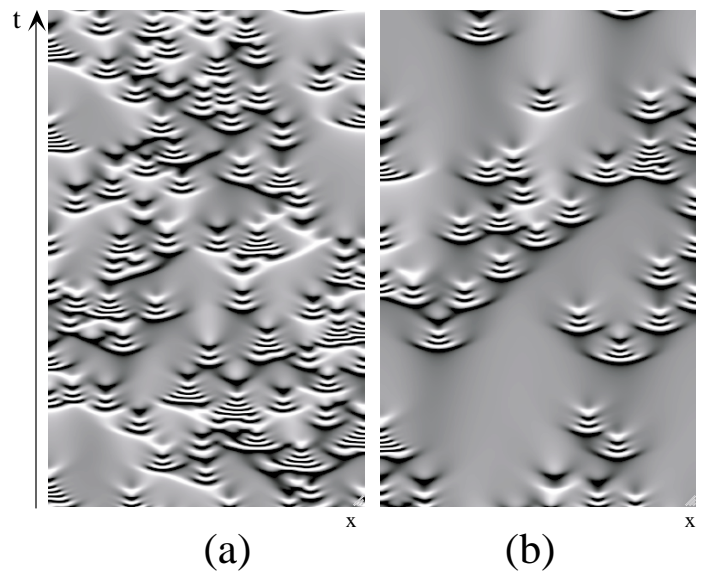

FIG. 1: $x - t$ diagram of Eq. (1) for $\alpha = 3.8$ and (a) $\lambda = 0.1$ (b) $lambda = 0.01$ (numerical scheme is a finite differences in space with 400 grid points with a spatial step 0.2, Runge-Kutta in time with time step 0.005)

attractor. The dynamics depends upon the history of the system. In a typical numerical experiment $\lambda$ is increased from negative to positive values. Above the pitchfork bifurcation, a transition to an inhomogeneous state occurs (see Fig. (1)). The system attempts to reach locally the large amplitude limit cycle. Strong phases gradients appear at the border between the domains of large oscillations and the ones of small amplitude. These gradients lead to fronts which propagate from the unstable zero state into the large amplitude oscillation and eliminate such patches. In this process large domains of the unstable non-oscillating state are created which in turn will evolve to new patches of oscillatory domains and fronts. This process is at the origin of a very interesting complex spatio-temporal behavior. It exhibits, for slightly positive values of $\lambda$, very strong intermittent features. A typical snapshot shows small patches of large amplitude oscillation embedded into large domains of the weakly unstable zero state.

The transition which takes place at $\lambda = 0$ is a very simple intermittent transition toward spatio-temporal complex behavior. The transition is characterized by critical exponents. The characteristic time diverges as $1/\lambda$ as is the case for type II intermittency in low-dimensional dynamical systems [3], and the characteristic domain size diverges as $1/\lambda$. The absence of hysteresis, the intermittent character of the transition, and the very nature of the complex behavior are related to the existence of fronts which "convectively clean" the small perturbations of the unstable zero state. These fronts will be called "retracting fronts" (RFs) in the following. They also exist in the subcritical parameter regime ($\lambda < 0$) where they are stable and connect the stable zero state to the large amplitude oscillatory state. In particular, when $\lambda < -0.25$, the topology of the phase portrait is very simple. The two attractors are the trivial stationary solution and the large amplitude limit cycle. They are separated by an unstable limit cycle which acts as a separatrix. The analysis of these fronts in the frame of Eq. (1) is not feasible.

To gain a better understanding simple amplitude equations, are more appropriate. The following modification of the dissipation function $\nu(u) = \epsilon - \alpha u^2 + u^4$ and of the potential $V = -\lambda u^2/2 - \eta u^4/2 + u^4/4$, allows, for negative $\lambda$, $\eta \sim \alpha \sim \epsilon$ and small $\epsilon$ to reduce Eq. (1) to the quintic complex Ginzburg-Landau equation (CGLe)

$$\partial_t A = \mu A + (1 + i\beta)\partial_x^2 A$$
$$+ (1 + ic_3)|A|^2 A + (-1 + ic_5)|A|^4 A \quad (2)$$

In the reduced units used there remain three system parameters (in addition to the control parameter $\mu$), namely $\beta$, $c_3$, and $c_5$, describing linear and nonlinear dispersion, respectively. Stationary fronts are solutions of the form $A = e^{-i\omega} B(\xi)$, $\xi = x - vt$. Writing $B = ae^{i\theta}$ with real $a$ and $\theta$ and introducing $q = \theta'$ the CGLe reduces to

$$(1+\beta^2)a'' + va' + R(a,q) + \beta I(a,q) = 0 \quad (3)$$
$$(1+\beta^2)(2qa' + q'a) - \beta(va' + R(a,q)) + I(a,q) = 0 \quad (4)$$
$$R(a,q) = (\mu - q^2 + a^2 - a^4)a$$
$$I(a,q) = (\omega + vq - \beta q^2 + c_3 a^2 + c_5 a^4)a \quad (5)$$

which corresponds to a 3−dimensional dynamical system. Plane-wave solutions of Eq. (2) correspond in Eq. (3,4) to constant solutions with wave number $q = q_N$. Then $a = a_N$ and $\omega = \omega_N$ are given by

$$a_N^2 = \frac{1}{2} \pm \sqrt{\frac{1}{4} + \mu - q_N^2} \quad (6)$$
$$\omega_N = -vq_N + \beta q_N^2 - c_3 a_N^2 - c_5 a_N^4$$
$$= -vq_N + \beta\mu - C_3 a_N^2 - C_5 a_N^4. \quad (7)$$

where $C_3 = c_3 - \beta$, $C_5 = c_5 + \beta$. They bifurcate at $\mu = q_N^2$ subcritically from the trivial state. At $\mu = q_N^2 - 1/4$ there is a saddle-node bifurcation leading to an amplitude-stable nonlinear branch corresponding to the upper sign in (6). One can easily show that the band-center solution $q_N = 0$ is modulationally stable for $D_0 = [1 - \beta(c_3 + 2c_5)]a_N^2 + 2(1 - \beta c_5)\mu > 0$. When $C_3 = C_5 = 0$ all plane waves have in the rest frame the same frequency and the group velocity $d\omega_N/dq_N$ is zero.

Front solutions with $a \to 0$, $q \to q_L$ on one side (we will here choose this to be at $\xi \to -\infty$) and $a \to a_N$, $q \to q_N$ on the other side ($\xi \to +\infty$), and thus $\omega = \omega_N$, have been studied in particular by van Saarloos and Hohenberg [4]. We are interested in nonlinear fronts whose velocity is determined from Eqs. (3) and (4) as a nonlinear eigenvalue. There exists a class of analytic ("polynomial") fronts of the form

$$q = q_N + e_0(a^2 - a_N^2), \quad a' = e_1 a(a^2 - a_N^2) \quad (8)$$

which select a velocity $v^+$ and a plane wave $q_N^+$. For the real case $\beta = c_3 = c_5 = 0$ one has $q_N^+ = 0$ and $v^+ = 4(1/4 - \sqrt{\mu + 1/4})/\sqrt{3}$. Then there exists a potential $U(a_N) = \frac{a_N^4}{3}(\sqrt{\mu + 1/4} - 1/4)$ and the front moves in such a direction that the potential decreases. The Maxwell point, where the front velocity reverses, is at $\mu_M = -3/16$.

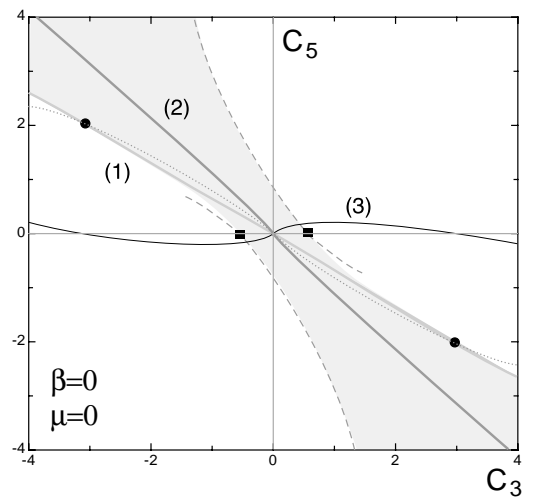

FIG. 2: Range of existence of fronts in the $c_3, c_5$ plane for $\mu = \beta = 0$. RFs exist outside the narrow region enclosed by curve 1, where their velocity becomes zero, and curve 2. Above curve 3 (for $c_3 + c_5 > 0$, otherwise below) their velocity increases with increasing $\mu$. The range of stable existence of NFs (approximately the shaded region) is limited by different criteria. Above the black square (for $c_3 + c_5 > 0$, otherwise below) the limit of amplitude stability of the plane wave is reached. Below the black square the range is first limited by the disappearance of the solution (dashed curve) and then by the group velocity changing sign. On the dotted curve the two types of fronts are glued together for an appropriate value of $\mu$ ($\mu = 0$: thick dots).

The selected wavenumber can invade the system only when the group velocity $\tilde{v}_g = d\omega_N/dq_N - v^+$ of the emitted plane wave in the rest frame of the front is positive, i.e. when the fronts act as sources. This is indeed the case in the main part of the region where polynomial front solutions exist, and here we will call them "normal fronts" (NFs). The range of (stable) existence of NFs is shown in Fig. 2 in the $c_3, c_5$ plane for $\mu = 0$, $\beta = 0$ (shaded region; since there is a symmetry $A$, $c_3$, $c_5$, $\beta \to A*$, $-c_3$, $-c_5$, $-\beta$ we need to discuss only the half plane $c_3 + c_5 > 0$). The region is limited either by reaching the saddle-node of the plane wave where the solutions become amplitude unstable, by reaching the limit where the solutions cease to exist (it merges in a saddle-node bifurcation with a faster, unstable front), or by $\tilde{v}_g$ becoming negative (for details see figure caption). The velocity $v^+$ decreases with increasing $\mu$.

From simulations of Eq. (2) we observed that stable RFs, which are not of the form (8), exist essentially everywhere outside the shaded region of Fig. 2, and also between the dashed curve and the solid curve 2. An example will be shown below. They represent sinks in their rest frame and do not select the wavenumber $q_N$. Thus there is a whole family of such front solutions, in particular the one with $q_N = 0$, which we have studied. RFs do not change their character when $\mu$ becomes positive, where the trivial state generated is linearly unstable.

The RFs (with $q_N = 0$) connect smoothly to the NFs when $\mu$ is decreased only on a hyperplane $\mathcal{H}_M$ in $c_3$, $c_5$, $\beta$ parameter space. $\mathcal{H}_M$ is obtained by requiring $v^+ = 0$ and $q_N^+ = 0$ in the relations given in [4] for the NFs. The transition occurs when $\mu$ crosses $\mu_M = a_M^2(1 - a_M^2)$ where $a_M^2 = -2C_3/(3C_5)$ (for $(C_3 C_5 \neq 0)$. $\mathcal{H}_M$ reduces to $c_5 = -16c_3/(15 + \sqrt{9 + 8c_3^2})$ for $\beta = 0$ (dotted line in Fig. 2). Then $\mu_M = 0$ at $C_3 = \pm 3$, $C_5 = \mp 2$ (thick dots). The transition at $\mathcal{H}_M$ can be viewed as a generalization of the Maxwell point of the potential case. The situation differs by the facts that for $\mu > \mu_M$ a nonzero $q_N$ is in general selected and for $\mu < \mu_M$ the solution cannot be obtained analytically. This holds also for $C_3 = C_5 = 0$ where one has $\mu_M = -3/16$, as in the potential case, which is indeed the lowest possible value.

Away from $\mathcal{H}_M$ the transition between NFs and RFs is discontinuous (at least slightly; actually there are small regions where neither front exists stably). In a large parameter range, the velocity of RFs increases with increasing $\mu$ (above the curve (3) in Fig. 2), which is counterintuitive and opposite to the behavior of normal fronts.

RFs are readily obtained numerically from the ODEs Eqs. (3) and (4) by shooting from the linear asymptotic regime $a \sim a_0 e^{\kappa_L \xi}$, $q \sim q_L$ for $\xi \to -\infty$ with $\kappa_L (> 0)$ and $q_L$ determined from the corresponding eigenvalue problem. Connection to the nonlinear state for $\xi \to +\infty$ involves, for $v > 0$, two stable and one unstable direction. Adjusting the velocity allows to eliminate the unstable direction, and thus to complete the heteroclinic orbit. The eigenvalues corresponding the stable directions can be real or imaginary. In the latter case, which occurs only for $\beta \neq 0$, the front is non-monotonic at the nonlinear side.

RFs can be captured perturbatively for $\mu > 0$ in the limit $C_3$, $C_5 \to 0$ where $v \to 0$ and spatial variations become slow. At lowest order (subscript 0) one obtains from (3,4) $R(a, q_0)) = I(a, q_0) = 0$. From this follows $q_0(a) = \sqrt{\mu + a^2(1-a^2)}$ $(> 0)$ and $C_3 = 0$, $C_5 = 0$, i.e. the front connects locally plane-wave states, which on this line in parameter space all have the same frequency, see (7). At next order we obtain from (4) $(1+\beta^2)(2q_0 a' + q_0' a) + I(a, q_0) = 0$ which can be rewritten as

$$D_a a' = (\omega + vq_0(a) - \beta \mu + C_3 a^2 + C_5 a^4), \quad (9)$$
$$D_a = (1+\beta^2)(a^2(4a^2 - 3) - 2\mu)/q_0(a). \quad (10)$$

Eq. (9) is a generalization of the stationary version of the usual phase equation (transform $a'$ into $q_0' = \theta''$). $D_a$ is related to the phase diffusion constant and for $D_a > 0$ plane-wave solutions are modulationally stable.

Along a front the local plane waves necessarily cross into the amplitude unstable regime. Such solutions of (9) can exist only when the zero of $D_a$ at $a_c^2 = (3 + \sqrt{9 + 32\mu})/8$ is compensated by a zero of the right-hand side. Taking $\omega = \omega_N$ from Eq. (7) fixes the velocity, which for $q_N = 0$ becomes

$$v = [c_3 - \beta + (c_5 + \beta)(a_N^2 + a_c^2)]\sqrt{\frac{a_N^2 - a_c^2}{a_N^2 + a_c^2 - 1}} \quad (11)$$

We have verified this result numerically. It gives a reasonable estimate for $v$ even when $C_3$ and/or $C_5$ become about 1 and also when $\mu$ is slightly negative. For $\mu < 0$ the solutions of Eq. (9) approximate the RFs well only for not too small values of $a$, i.e. outside the tail region. The slopes of the various curves at the origin of Fig. 2 can now be calculated.

We now come to the qualitative understanding of the nature of the new fronts. The mechanism leading to the RFs motion can be understood by multiplying Eq. (3) by $a'$ and integrating over $\xi$. Writing $\int_{-\infty}^{+\infty} ... = <...>$ one obtains for $\beta = 0$

$$v<a'^2> = -U(a_N) + <q^2 aa'> \quad (12)$$

where $U(a_N) = \frac{a_N^4}{3}(\sqrt{\mu + 1/4} - 1/4)$ is the "potential term" pushing the front. The last term arises only in the nonpotential case when a phase gradient develops, as is seen from Eq. (4), which can now be written in the form of a "spatial transport equation" $(a^2 q)' + v(a^2 q) = -(\omega + c_3 a^2 + c_5 a^4) a^2$. Clearly, the phase gradient term in Eq. (12) drives the front towards the nonlinear state.

As an attempt to further reduce the problem, we have also considered the peculiar limit of Eq. (2) which consists in keeping only the cubic nonlinear term and the diffusion term.

$$\partial_t A = ic_3|A|^2 A + A_{xx} \quad (13)$$

A "toy physical model" associated with this equation is a chain of pendula coupled by viscous torques. The "nonlinear complex diffusion" equation (13) possesses a continuum of homogeneous oscillating solutions $A = A_0 \exp ic_3|A_0|^2 t$ related by a scaling transformation. Consider as an initial condition a sharp front between the state $A = 0$ and a homogeneous oscillatory state. The effect of the diffusion is to blur the interface while the effect of the nonlinear term is to create a phase gradient localized at its core (see Fig. (3)). This localized phase gradient "pushes" the front in the direction of the homogeneneous oscillating state, since the highest gradients, which dissipate the most, develop on the side of the sub-critical state. Asymptotically a front forms. Whereas a soliton of the focusing nonlinear Schrödinger equation balances nonlinearity and dissipation, our front balances nonlinearity and diffusion. The nonlinear oscillations transform dissipation into front translation. The width of the front is of the order $1/(|c_3||A_0|)$, and its velocity $v \sim |c_3||A_0|$.

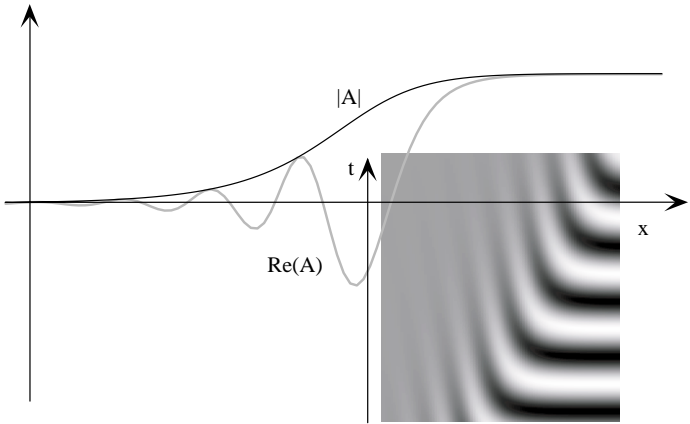

FIG. 3: Front solution of Eq. (13) for $c_3 = -1$ (numerical scheme is a finite differences in space with 400 grid points, Runge-Kutta in time with time step 0.01). The real part and the modulus part of $A$ are shown. The "x-t" diagram illustrate the motion of the front.

We mention that by retaining in Eq. (13) also the term $ic_5|A|^4 A$ one can, for $c_3 c_5 < 0$, find analytic (polynomial) RFs for a selected frequency out of the continuum of oscillating states. They generalize the polynomial fronts of ref. [4], which for $q_N = 0$ exist only for $c_3 = -2c_5$.

Clearly, RFs were at the origin of the intermittent transition to complex behavior in Eq. (1). Although the amplitude equation (2) was mainly used as a tool to get analytic handles on these solutions, it was tempting to study numerically the transition when $\mu$ crosses zero in this model too. As we could have expected, the transition has the same intermittent features (see Fig. (4)). Numerical calculation of the mean domain size show that it behave as $1/\mu$ (we have tested this for $c_3 = c_5 = 2$, $\beta = 0.4$). Thus the intermittent transition mediated by RFs appears as a generic mechanism to complex behavior. We mention that the absence of hysteris is by no means a general feature in the transition to spatio-temporal complexity. Indeed, in the CGLe in the parameter regime where NFs and RFs coexist, one can observe substantial hysteresis.

Localized states (pulses), which have received much attention in subcritical bifurcations [5–7] can be interpreted to a large part as bound states of two RFs. Their stable existence is due to the repulsive interaction of two fronts. We found that also different types of hole-like localized states, which in the CGLe connect nonlinear states with $q_N = 0$, exist and can be understood in terms of bound RFs. Finally we mention that in the large parameter space of the CGLe RFs can become unsta-ble. When $\beta$ becomes sufficiently large one finds oscillating and chaotic fronts.

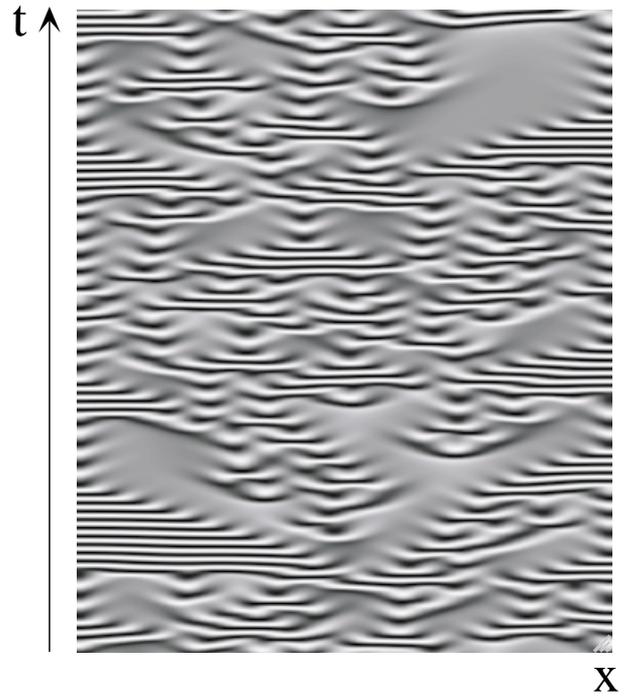

FIG. 4: $x - t$ diagram of Eq. (2) for $\mu = 0.1$, $c_3 = .9$, $c_5 = \beta = 0$

In conclusion: we have found a new class of front solutions in nonpotential systems which mediates a multitude of striking phenomena many of which are of experimental interest. It now appears important to model experiments like that of ref. [2] more realsitically.

This work was performed to a large part while one of us (PC) was visiting the Physics Institute of the University of Bayreuth, supported by an Alexander-von-Humboldt award. The simulations were done with the XDim Interactive Simulation Package developed by one of us (PC) and M. Monticelli, who we thank for his support.